\documentclass[epj]{svjour}
\usepackage{graphicx}
\begin{document}

\input epsf.sty

\title{Bond dilution in the 3D Ising model: a Monte Carlo study}
\titlerunning{Bond dilution in the 3D Ising model}

\author{Pierre-Emmanuel Berche\inst{1}\thanks{Author for correspondence
(pierre.berche@univ-rouen.fr).}, Christophe Chatelain\inst{2},
Bertrand Berche\inst{2} \and Wolfhard Janke\inst{3}}
\authorrunning{P.E. Berche, C. Chatelain, B. Berche and W. Janke}

\institute{Groupe de Physique des Mat\'eriaux (UMR CNRS No 6634),
Universit\'e de Rouen,\\
F-76801 Saint Etienne du Rouvray Cedex, France
\and
Laboratoire de Physique des Mat\'eriaux (UMR CNRS No 7556),
Universit\'e Henri Poincar\'e, Nancy 1,\\
F-54506 Vand\oe uvre les Nancy Cedex, France
\and
        Institut f\"ur Theoretische Physik, Universit\"at Leipzig,
Augustusplatz 10/11,\\
D-04109 Leipzig, Germany
}

\date{\today}

\abstract{
We study by Monte Carlo simulations the influence
of bond dilution on the three-dimensional
Ising model. This paradigmatic model in its pure version displays a second-order phase
transition with a positive specific heat critical exponent $\alpha$. According to the Harris criterion
disorder should hence lead to a new fixed point characterized by new critical exponents. We have
determined the phase diagram of the diluted model, between the pure model limit and the percolation threshold.
For the estimation of critical exponents, we have first performed a finite-size scaling study, where we
concentrated on three different dilutions. We emphasize in this work the great influence of the cross-over
phenomena between the pure, disorder
and percolation fixed points which lead to effective critical exponents dependent on the concentration.
In a second set of simulations, the temperature behaviour of physical quantities has been studied in
order to characterize the disorder fixed point more accurately. In particular this allowed us to estimate
ratios of some critical amplitudes.  In accord with previous observations for other models this provides
stronger evidence for the existence of the disorder fixed point since the amplitude ratios are 
more sensitive to the universality class than the critical exponents. Moreover, the question
of non-self-averaging at the disorder fixed point is
investigated and compared with recent results for the bond-diluted $q=4$ Potts model. Overall our numerical
results provide evidence that, as expected on theoretical grounds, the critical behaviour of the bond-diluted
model is governed by the same universality class as the site-diluted model.
}

\maketitle

\newcommand{\centre}[2]{\multicolumn{#1}{c}{#2}}
\newcommand{\crule}[1]{\multispan{#1}{\hrulefill}}
\def\br{\noalign{\vskip2pt\hrule height1pt\vskip2pt}}


\section{Introduction}\label{s1}

The influence of quenched, random disorder on phase transitions has been the
subject of numerous experimental and theoretical investigations since more
than 20 years~\cite{cardy96}.
They concern both first- and second-order phase transitions, especially in
two dimensions. The qualitative influence of quenched, short-range
correlated~\cite{Note} random disorder at
second-order phase transitions is well understood since Harris~\cite{Harris74}
proposed a relevance criterion based on the knowledge of the specific heat
critical exponent $\alpha$ of the pure model: when $\alpha$ is positive,
under a coarse graining, the
disordered system should reach a so-called finite-randomness disorder fixed
point characterized
by altered critical exponents, whereas if $\alpha$ is negative, the
universality class of the pure system will persist. In particular the
two-dimensional (2D) Ising model has attracted great interest in the past years
because of its intermediate situation ($\alpha=0$) which implies a marginal
influence of disorder~\cite{Shalaev94}. From an experimental point of view, a
confirmation of the Harris criterion was reported in a LEED investigation of
a 2D order-disorder transition~\cite{SchwengerEtAl94,VogesEtAl98}. In the case
of a first-order phase transition in the pure model, disorder is expected to
soften the transition and, under some circumstances, may even induce
a second-order transition~\cite{ImryWortis79}. For 2D systems, the latter
scenario has been proved by Aizenman and Wehr on rigorous theoretical
grounds~\cite{AizenmanEtAl89,HuiEtAl89}. To test these theoretical predictions
also the Potts model~\cite{Wu82} has been intensively studied
in 2D~\cite{BercheChatelain03} since it
displays the two different regimes: a second-order phase
transition when the number of states per spin $q\leq 4$ and a first-order one
when $q>4$. These results were obtained by different techniques including
Monte Carlo simulations, transfer matrix calculations, field-theoretic
perturbation theory, and high-temperature series expansions
~\cite{Ludwig87,Chen92,Dotsenko95,Jug96,Cardy97,Picco97,%
Jacobsen98,Chatelain98,Roder98,Olson99,Chatelain00}.

In three dimensions (3D), the disordered Potts model has of course been studied
only later: the
case $q=3$, corresponding to a very weak first-order transition for the pure
 model~\cite{JaVi97}, has been investigated numerically by Ballesteros
{\em et al.\/}~\cite{BallesterosEtAl00} for site dilution, and the case $q=4$,
exhibiting in the pure system a strong first-order
transition~\cite{JankeKappler96}, has been studied very recently by
us~\cite{Chatelain01,Chatelain02,Chatelain02bis,Berche02bis,athens03} for
bond dilution via large-scale Monte Carlo simulations.
Only the 3D Ising model with site dilution has also been extensively studied
using Monte Carlo simulations~\cite{Landau80,Marro86,Chowdhury86,Braun88,%
Wang89,Wang90,Holey90,Heuer90a,Heuer90b,Heuer93,Hennecke93,%
Ballesteros98,Wiseman98a,Wiseman98b,CalabreseEtAl03Bis},
or field theoretic renormalization group approaches~\cite{Holovatch98,Folk98,%
Folk99,Folk00,Folk01,Varnashev00,Pakhnin00,Pakhnin00b,Pelissetto00,%
CalabreseEtAl03ter}.
The diluted model can be treated in the low-dilution regime (concentration of
magnetic bonds, $p$, close to $1$) by analytical perturbative renormalization
group methods~\cite{Newman82,Jug83,Mayer89,Mayer89b,Bervillier92} where a
new fixed point independent of the dilution has been found, but for stronger
disorder only Monte Carlo simulations remain valid.

The first numerical studies~\cite{Marro86,Chowdhury86,Braun88} suggested a
continuous variation of the critical exponents along the critical line but
after the works~\cite{Wang89,Wang90,Holey90,Heuer90a,Heuer90b} it became clear that
the concentration dependent critical exponents found in Monte
Carlo simulations are effective ones, characterizing the approach to the
asymptotic regime. The critical exponents $\beta$ and $\gamma$ associated with
the magnetisation and susceptibility, respectively, were shown by
Heuer~\cite{Heuer90a} to be concentration dependent in the region
$0.5\leq p< 1$. The conclusion of Heuer was: while a crossover between the pure
and weakly random fixed points accounts for the behaviour of systems above
$p\simeq 0.8$, in more strongly disordered systems a more refined analysis is
needed. We should mention here that the meaning of $p$ is not exactly the same
in the papers mentioned above which refer to site dilution and in our present
study in which we are interested in bond dilution. That is why the numerical
values of $p$ given before for the different regimes cannot directly be
taken over to the bond-dilution case.

Another important question has been investigated by Wiseman and
Domany~\cite{Wiseman98a,Wiseman98b}: it concerns the question of the possible
lack of self-averaging which can happen in disordered systems. For the 3D
site-diluted Ising model close to criticality they explicitly showed that
physical quantities such as the magnetisation or susceptibility are indeed not
self-averaging. Although simulations~\cite{Wiseman98b} revealed that
disorder realized in a canonical manner (fixing the fraction $p$ of magnetic
sites) leads to different results than those obtained from disorder realized
in a grand-canonical ensemble (assigning to each site a magnetic moment with
probability $p$), the renormalization-group approach of Ref.~\cite{Aharony98}
shows that the canonical
constraint is irrelevant, even near the random fixed point, suggesting that the observed
differences are a finite-size effect.
The studies of Ref.~\cite{Ballesteros98} were based on the
crucial observation that it is important to take into account the leading
corrections-to-scaling term in the infinite-volume extrapolation of the Monte
Carlo data. Thus, the main problem encountered in these studies of the
disordered Ising model was the question of measuring effective or asymptotic
exponents. Although the change of universality class should happen
theoretically for arbitrarily weak disorder, the new critical exponents
appear only in a small temperature region around the critical point, whose size
is controlled by the concentration of the non-magnetic compound. Equivalently,
in finite-size scaling studies very large lattices are required to observe the
asymptotic behaviour. In fact, the asymptotic regions cannot always be reached
practically and one therefore often measures only effective exponents.

Another crucial problem of the new critical (effective or asymptotic) exponents
obtained in these studies is that the ratios $\beta/\nu$ and $\gamma/\nu$
occuring generically in finite-size scaling analyses are almost identical
for the disordered and pure models. For the pure 3D Ising model, accurate values
are~\cite{Guida98}:
$$\nu=0.6304(13),\quad\eta=0.0335(25),$$
which gives:
$$\beta/\nu=0.517(3), \quad \gamma/\nu=2-\eta=1.966(3),$$
and $\alpha=0.1103(1) > 0$, i.e., disorder should be relevant according to
the Harris criterion.

For the site-diluted model, the asymptotic exponents given by Ballesteros
{\em et al.}~\cite{Ballesteros98} are:

$$\beta/\nu=0.519(3), \quad \gamma/\nu=1.963(5), \quad \nu=0.6837(53).$$

Thus, finite-size scaling techniques will only be able to differentiate between
the values of $\nu$ at the two fixed points, but will not be very efficient
for distinguishing ratios of critical exponents.
Even if $\beta$ and $\gamma$ themselves happen to be quite different, the
ratios $\beta/\nu$ and $\gamma/\nu$ are very close. That is the reason why a
study of the temperature behaviour of the magnetisation and susceptibility
will be very helpful for an independent determination of the exponents
$\beta$ and $\gamma$.

Contrary to previous studies of the disordered Ising model which were concerned
with site dilution, we have chosen to model the disorder by bond dilution in
order to compare these two kinds of disorder and to verify that they indeed lead
to the same set of new critical exponents, as expected theoretically by
universality arguments. In the following we shall thus consider the
bond-diluted Ising model in 3D whose Hamiltonian with uncorrelated quenched
random interactions can be written (in a Potts model normalization) as
\begin{equation}
-\beta {\cal H}=\sum_{(i,j)}K_{ij}\delta_{\sigma_i,\sigma_j},
\end{equation}
where the spins take the values $\sigma_i=\pm 1$ and the sum goes over
all nearest-neighbour pairs $(i,j)$. The coupling strengths are allowed to take
(grand-canonically) two different values $K_{ij}= K \equiv J/k_BT$ and $0$
with probabilities $p$ and $1-p$, respectively,
\begin{eqnarray}
{\cal P}[K_{ij}]&=&\prod_{(i,j)}P(K_{ij})\nonumber\\
&=&\prod_{(i,j)}[p\delta(K_{ij}-K)+(1-p)\delta(K_{ij})],
\end{eqnarray}
$c=1-p$ being the concentration of missing bonds, which play the role of the
non-magnetic impurities.

The plan of the rest of the paper is as follows: in section 2, we present the
phase diagram of the bond-diluted Ising model. Then, in section 3,
we discuss the averaging procedure and investigate the question of possible
non-self-averaging of physical quantities. The section 4 is devoted to the
critical behaviour of the disordered model. In the first part we present our
finite-size scaling study of three particular concentrations $p$, and the
second part deals with the temperature scaling for the same three dilutions.
Finally, section 5 contains our conclusions.

\section{Phase diagram}
\label{s2}

In order to determine the phase diagram and the critical properties at a few
selected dilutions we performed in this study large-scale Monte Carlo
simulations on simple cubic lattices with $V=L^3$ spins (up to $L=96$) and
periodic boundary conditions in the three space directions, using the
Swendsen-Wang cluster algorithm~\cite{Swendsen87} for updating the spins.
The histogram reweighting technique~\cite{Ferrenberg88,Ferrenberg89} was
employed to extend the results over a range of $K$ around the simulation
point. All physical quantities are averaged over 2\,000 -- 5\,000 disorder
realisations, indicated by a bar (e.g., $\bar m$ for the magnetisation).
Standard definitions were used, e.g., for a given disorder realisation,
the magnetisation is defined according to $m=\langle|\mu|\rangle$ where
$\langle\dots\rangle$ stands for the thermal average and
$\mu=(N_\uparrow-N_\downarrow)/(N_\uparrow+N_\downarrow)$. The susceptibility
follows from the fluctuation-dissipation relation,
$\chi=KV(\langle\mu^2\rangle-\langle|\mu|\rangle^2)$. Here, one should notice that
a definition which is usually used in the disordered phase in Monte Carlo
simulations,
and which seems to be more stable, $\chi=KV\langle\mu^2\rangle$, should
be avoided in quenched disordered systems. Simulating at a given temperature
above the critical point, one  may indeed encounter samples which have
higher effective transition temperatures, and which thus have non-zero
$\langle|\mu|\rangle$.

\begin{figure}[b]
\begin{center}
\resizebox{0.80\columnwidth}{!}{%
  \includegraphics{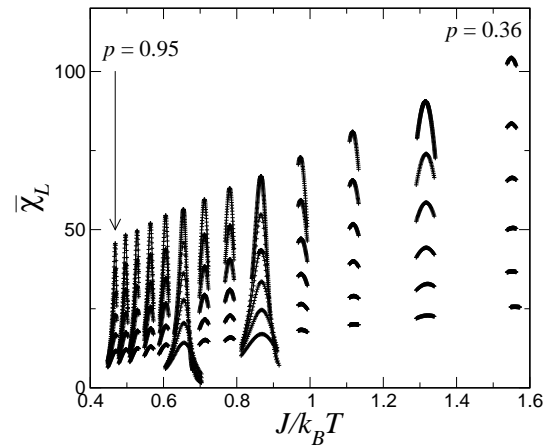}}
\end{center}\vskip 0cm
\caption{Variation of the average magnetic susceptibility $\bar\chi_L$ versus the coupling
strength $K=J/k_B T$ for several concentrations $p=0.95, 0.90, \dots, 0.36$
and $L=10,12,14,16,18,20$. For each value of $p$
and each lattice size, the curves are obtained by standard histogram reweighting
of the simulation data at
one value of $K$.}
\label{Chi}
\end{figure}

The phase diagram was obtained by locating the maxima of the average
susceptibility $\bar\chi_L$ (a diverging quantity in the thermodynamic limit)
for increasing lattice sizes $L$ as a function of the coupling strength $K$,
with the
dilution parameter $p$ varying from the neighbourhood of the pure model
($p=0.95$) to the very diluted model ($p=0.36$), see Fig.~\ref{Chi}.

\begin{figure}[t]
\begin{center}
        \resizebox{0.80\columnwidth}{!}{%
        \includegraphics{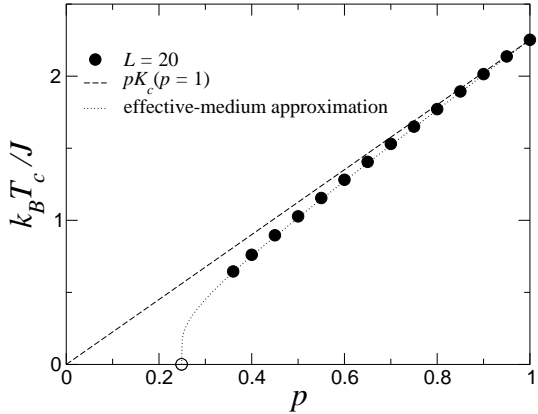}}
\end{center}\vskip 0cm
\caption{Phase diagram of the 3D bond-diluted Ising model compared
                 with the mean-field and effective-medium approximations.
The open circle marks the location of the percolation threshold.}
\label{0-Kc-vs-p}
\end{figure}

Below the percolation threshold $p_c\simeq 0.2488$~\cite{LorenzZiff98}, one
does not expect any finite-temperature phase transition since without any
percolating cluster in the system long-range order is impossible. The
approximate phase diagram~\cite{Berche02} as obtained from the susceptibility
maxima for the largest lattice size $(L=20)$ is shown in Fig.~\ref{0-Kc-vs-p}.
For comparison, we have drawn a simple mean-field (MF) estimate of the
transition point
\begin{equation}
K_c^{{\rm MF}}(p) = p K_c (1),
\end{equation}
where $K_c(1) = 0.443\,308\,8(6)$ \cite{talapov} is the accurately known
transition point of the pure model (in the Potts model normalization),
and the single-bond effective-medium (EM) approximation~\cite{Turban80}
\begin{equation}
K_c^{{\rm EM}}(p)=\ln\left [{(1-p_c)e^{K_c (1)}-(1-p)\over p-p_c}\right],
\end{equation}
which gives a very good agreement with the simulated transition line over the
full dilution range. We have omitted results from
recent high-temperature series
expansions \cite{meik_is} since, on the scale of Fig.~\ref{0-Kc-vs-p}, they
would just fall on top of the Monte Carlo data.

\begin{figure}[ht]
\begin{center}
        \resizebox{0.80\columnwidth}{!}{%
        \includegraphics{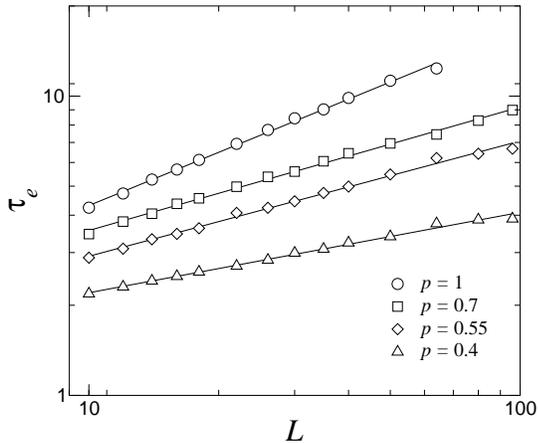}}
\end{center}\vskip 0cm
\caption{Energy autocorrelation time $\tau_e$
                 versus the size $L$ of the system on a log-log scale. The
                 pure case corresponds to $p=1$. The straight lines show fits
                 of the form $\tau_e \propto L^z$, yielding the
                 effective dynamical critical exponents $z$ 
		 compiled in Table~\ref{tab:table1}.}
\label{Tau-E}
\end{figure}

To get an accurate determination of $K_c(L)$, we used the histogram
reweighting technique with at least $N_{\rm MCS} = 2500$ Monte Carlo sweeps
and between $2000$ and $5000$ samples of disorder. The number of Monte Carlo
sweeps is justified by the increasing behaviour of the energy autocorrelation
time $\tau_e$ as a function of $p$ and $L$. For each size, we performed at least
$250$ independent measurements of the physical quantities
$(N_{\rm MCS} > 250\ \tau_e)$.
For a second-order phase transition, the autocorrelation time is expected to
behave as $L^z$ at the critical point where $z$ is the dynamical critical
exponent. For the disordered Ising model, we obtained from the least-squares
fits shown in Fig.~\ref{Tau-E} the values of $z$ compiled in
Table~\ref{tab:table1}.
We see that the critical slowing-down weakens for the disordered model and
that $z$
becomes 
effectively
smaller when the concentration of magnetic bonds $p$ decreases. The
observed
variation of the dynamical exponent $z$ is probably due to the influence of
the different fixed points encountered (pure, disorder and percolation), as 
will be discussed below in more detail for the static critical exponents.

The
largest autocorrelation time observed for the disordered model was around
$\tau_e\approx 9$ for $p=0.7$ and $L=96$.

\begin{table}
\small
\caption{Effective dynamical critical exponent $z$ as obtained from linear fits of
$\log \tau_e$ vs $\log L$. The simulations are performed at the
pseudo-critical couplings $K_c (L)$ (cf.\ Table~\ref{table_values_K} and
Fig.~\ref{Kc-vs-1surL} below).
Here we give in the second line only the simulation coupling $K$ for the
largest lattice size $L=96$.}
\vglue0mm\begin{center}
\begin{tabular}{@{}*{6}{l}}
\br
   $p$ & $1$ & $0.7$   & $0.55$  &  $0.4$   \\
$K$ & $0.4433$ & $0.6535$ & $0.8649$ & $1.3136$\\
\hline
$z$ & $0.59$ & $0.41$ & $0.38$ & $0.27$\\
\br
\end{tabular}
\\
\medskip\end{center}
\label{tab:table1}
\end{table}

\section{Non-self-averaging}
\label{s3}

\begin{figure}[t]
\begin{center}
        \resizebox{0.90\columnwidth}{!}{%
        \includegraphics{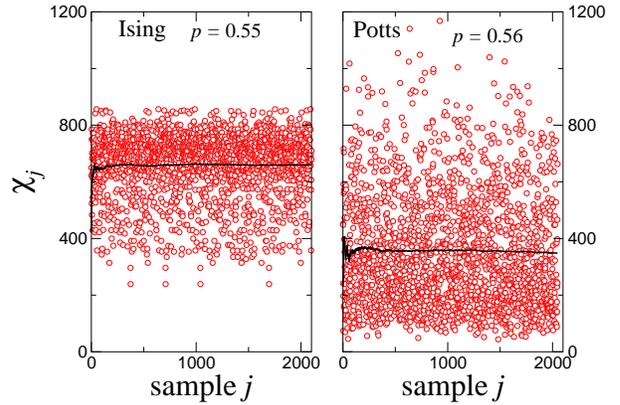}}
\end{center}\vskip 0cm
\caption{Disorder distribution of the susceptibility for
the Ising and $q=4$ Potts models with a concentration of magnetic bonds
of $0.55$ and $0.56$, respectively, and a lattice of size $L=64$. The
simulations are made at $K=0.8655 \approx K_c(L)$ for the Ising model and
$K=1.12945 \approx K_c(L)$ for the Potts model. In the latter case, the
concentration $p=0.56$ belongs to the second-order regime~\cite{Chatelain01}. The running average
over the samples $\bar\chi_j$ is shown by the thick solid line.}
\label{Convergence}
\end{figure}

In order to achieve accurate results for quenched, disordered systems in
numerical simulations it is important to obtain an estimate of the required
number of disorder realisations. This is particularly important in the vicinity
of a critical point where the correlation length diverges. As a
consequence the
(disorder) distributions of physical observables typically do not become
sharper with increasing system size at a finite-randomness disorder fixed point.
Rather their relative widths stay
constant, a phenomenon called non-self-averaging.
In order to investigate the disorder averages, we produced $N_s$ different
samples and computed the corresponding susceptibilities $\chi_j$,
$1\leq j\leq N_s$. In Fig.~\ref{Convergence} we compare the distributions
for the Ising and $q=4$ Potts models (in the second-order
regime~\cite{Chatelain01}).
The figure shows that the dispersion of the values of
$\chi$ is less important for the disordered
Ising model where rare events correspond to low values of $\chi$. This
implies that their contribution to the average is not so crucial as for the
Potts model where a long rare-event tail is found on the large-$\chi$ side.
This is the reason why for the Ising model the fluctuations
in the average value disappear after a few hundred realisations and why the
disorder averaging procedure is more efficient than for the Potts model.

\begin{figure}[ht]
\begin{center}
        \resizebox{0.90\columnwidth}{!}{%
        \includegraphics{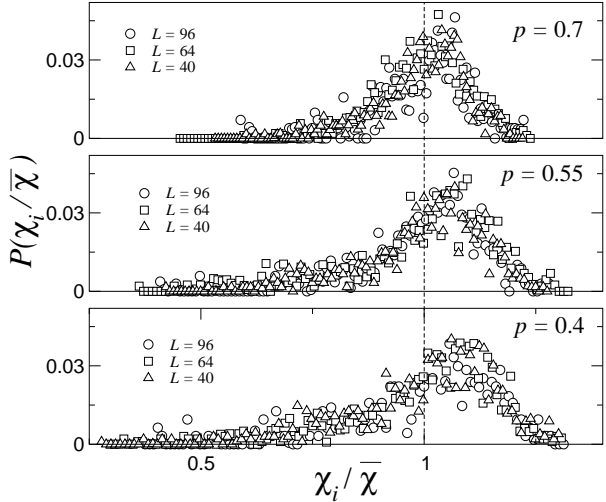}}
\end{center}\vskip 0cm
\caption{Probability distribution of the susceptibility versus the
                 ratio of the average value of the susceptibility of bin
                 $i$ and the global average susceptibility $\bar\chi$ for
                 the bond-diluted Ising model ($p=0.7$, $0.55$, and $0.4$) for
                 $L=40,64$, and $96$. The simulations are performed at
                 $K_c (L)$. The vertical dashed line indicates the average
                 susceptibility $\chi_i/\bar\chi=1$.}
\label{Proba-chi-IM}
\end{figure}

To test if self-averaging is valid or not for the disordered
model~\cite{Wiseman98a,Wiseman98b}, we investigated the variation of the shape
of the probability distributions of the susceptibility when the size
is increasing. Dividing the interval $(\chi_{\rm max}-\chi_{\rm min})$
into $100$ bins, we computed the probability for each value of $\chi$ to
belong to the bin $i$ ($1\leq i\leq 100$).
The resulting probability distribution $P(\chi_i/\bar\chi)$ (normalized to
unity) is drawn in Fig.~\ref{Proba-chi-IM} versus the ratio of the average
susceptibility $\chi_i$ of the bin $i$ and the global average susceptibility
$\bar\chi$.

\begin{figure}[t]
\begin{center}
        \resizebox{0.80\columnwidth}{!}{%
        \includegraphics{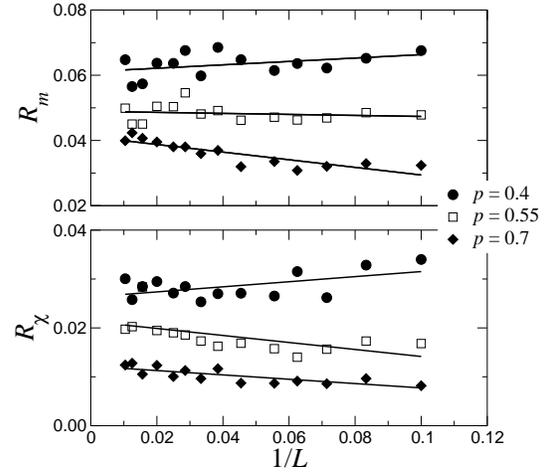}}
\end{center}\vskip 0cm
\caption{Normalized squared width of the magnetisation and susceptibility distributions
         versus the
                 inverse of the lattice size for the three concentrations
                 $p=0.4,0.55$, and $0.7$ at the effective critical coupling
                 $K_c (L)$. The straight lines are linear fits used as
guides to the eye.}
\label{Rapport-WD}
\end{figure}

For the three studied dilutions, the shape of the curve does not become
sharper with increasing size and the general shape remains remarkably
independent of the dilution, a phenomenon which strongly contrasts with the
situation encountered in the case of the $q\!=\!4$ Potts model~\cite{Chatelain01}.
Non self-averaging can be quantitatively checked by evaluating
the normalized squared width $R_\chi (L)=V_\chi (L)/\overline{\chi(L)}^2$,
where $V_\chi$ is the variance of the susceptibility
distribution: $V_\chi (L)= \overline{\chi^2(L)}-\overline{\chi(L)}^2$. The
same quantity is also evaluated for the magnetisation: 
$R_m (L)=V_m (L)/\overline{m(L)}^2$.
These
ratios are shown versus the inverse lattice size for the three studied
concentrations of the disordered Ising model in Fig.~\ref{Rapport-WD}.
The fact that $R_m$ and $R_\chi$ approach a constant when $L$ increases as 
predicted by
Aharony and Harris~\cite{Aharony96} is the signature of a non-self-averaging
system, in agreement with the results of Wiseman and
Domany~\cite{Wiseman98a,Wiseman98b} for the site-diluted 3D Ising model.
In fact, our numerical values of $R_m$ are close to the 
estimate $R_m = 0.055(2)$ for site dilution reported in Ref.~\cite{Wiseman98b}:
when the system size goes to infinity, our data are compatible with a convergence
towards $R_m = 0.05(1)$, where the error estimate reflects the uncertainty in 
the remaining finite-size corrections which appear to be more pronounced than for
site dilution. On the other hand, our limiting value for $R_\chi$ seems 
at first sight to be completely off from the estimate $R_{\chi'} = 0.156(4)$ 
given in Ref.~\cite{Wiseman98b}. Upon closer inspection this discrepancy
can be traced back to different definitions of the susceptibility. While
Wiseman and Domany worked with the ``high-temperature'' expression 
$\chi'=KV\langle\mu^2\rangle$, we used the connected correlator
$\chi=KV(\langle\mu^2\rangle-\langle|\mu|\rangle^2)$, valid in both the
low- and high-temper\-ature phase. Another difference is that they evaluated
$\chi'$ at the infinite-volume critical coupling $K_c^\infty$ while we followed the
pseudo-transition points $K_{\rm max}(L)$ defined from the location of the 
$\chi$ maxima (see below).
In fact, when we evaluate $R_{\chi'}$ (i.e., {\em without\/} subtraction), we
find for $p=0.7$ and the largest sizes (where $K_{\rm max}(L) \approx K_c^{\infty}$)
about 10 times larger values than shown in Fig.~\ref{Rapport-WD}:
$R_{\chi'} = 0.1635$, $0.1766$, and $0.1630$ for $L=64$, 80, and 96. This is
in accord with a remark in Ref.~\cite{Wiseman98b} who noticed also for
site dilution that $R_\chi$ is smaller by a factor of $~7-10$, which would
lead to a crass violation of the leading-order prediction in $\epsilon = 4-d$ 
for the ``ratio of ratios'' $R_m/R_\chi = 1/4$ ~\cite{Aharony96}, 
while working with $\chi'$ they obtained a much closer ratio of $R_m/R_{\chi'} = 0.35(2)$. 
In our $p=0.7$ case we get, with $R_m = 0.0407$, $0.0423$, and $R_m = 0.0399$ 
for $L=64$, 80, and 96, estimates of 
$R_m/R_{\chi'} = 0.2489$, $0.2395$, and $0.2448$, in perhaps surprisingly good
agreement with the renormalization group prediction of 1/4. Of course, given
the large corrections seen otherwise, this agreement may be accidental, but
it is also conceivable that the corrections do indeed partially cancel in the 
``ratio of ratios''.

Still, the apparent non-universality of $R_\chi$ and to a weaker
extent also of $R_m$ (the ratios seem
to depend on the impurity concentration) remains puzzling. It may be attributed to 
crossover effects between the pure,
disorder and percolation fixed points which may be more sensitive for bond dilution
than for site dilution.
Another possible reason for quantitative discrepancies might be the fact that 
$R_m$ and $R_\chi$ were estimated in our study
at the temperature corresponding to the maximum of the average susceptibility
for the corresponding lattice size, and not at the critical point of the
infinite system. This may lead to different scaling limits (similar to
the different Binder parameter scaling limits for pure systems in the
vicinity of $T_c$), albeit causing presumably only rather small deviations.

\section{Critical behaviour}\label{s4}

In order to study the critical behaviour of the bond-diluted Ising model,
we have concentrated on the three particular concentrations $p=0.7$, $0.55$,
and $0.4$, for which the simulated lattice sizes go up to $L=96$.
For interested readers, the values of the simulation temperatures are
reported in Table~\ref{table_values_K}.

\begin{table}
\small
\caption{Values of the simulation points $K=J/k_BT$ (extremely close
to the susceptibility maxima)
for the three mainly studied dilutions.}
\vglue0mm\begin{center}
\begin{tabular}{@{}*{5}{l}}
\br
$L$ & $p=0.70$ & $p=0.55$ & $p=0.40$ \\
\hline
10 & 0.6560 & 0.8680 & 1.3500 \\
12 & 0.6556 & 0.8680 & 1.3210 \\
14 & 0.6550 & 0.8675 & 1.3185 \\
16 & 0.6546 & 0.8655 & 1.3170 \\
18 & 0.6546 & 0.8655 & 1.3175 \\
22 & 0.6542 & 0.8655 & 1.3160 \\
26 & 0.6541 & 0.8650 & 1.3147 \\
30 & 0.6538 & 0.8650 & 1.3144 \\
35 & 0.6538 & 0.8650 & 1.3142 \\
40 & 0.6538 & 0.8655 & 1.3141 \\
50 & 0.6537 & 0.8653 & 1.3142 \\
64 & 0.6535 & 0.8655 & 1.3144 \\
80 & 0.6535 & 0.8649 & 1.3136 \\
96 & 0.6535 & 0.8649 & 1.3136 \\
\br
\end{tabular}
\\
\medskip\end{center}
\label{table_values_K}
\end{table}

\begin{figure}[t]
\begin{center}
        \resizebox{0.80\columnwidth}{!}{%
        \includegraphics{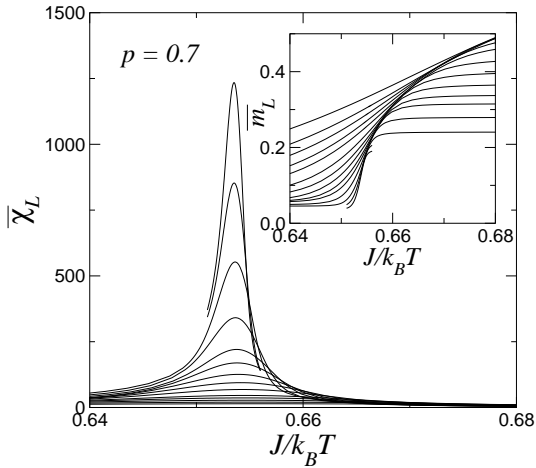}}
\end{center}\vskip 0cm
\caption{Variation of the magnetic susceptibility and of the
                 magnetisation (in the insert) versus the coupling strength
                 $K=J/k_B T$ for the bond-diluted Ising model ($p=0.7$) and
                 the lattice sizes compiled in Table~\ref{table_values_K} going
from 10, 12, 14, \dots to 96. For each value of $L$, the
                 curves are obtained by standard histogram reweighting of the
                 simulation data at the value of $K$ given in
Table~\ref{table_values_K}.}
        \label{chi-M-p=0.7}
\end{figure}

In Fig.~\ref{chi-M-p=0.7} we show results of typical runs for the
magnetisation (in the insert) and the susceptibility for $p=0.7$ at a
particular temperature (or coupling) very close to the size-dependent critical
point using the histogram reweighting technique. From the scaling of the
location and height of the peaks with lattice size of this and similar
quantities we then extracted the critical exponents of the system.

\subsection{Finite-Size Scaling Study}

\subsubsection{Correlation length exponent}

\begin{figure}[t]
\begin{center}
        \vskip -0.4cm
        \resizebox{0.90\columnwidth}{!}{%
        \includegraphics{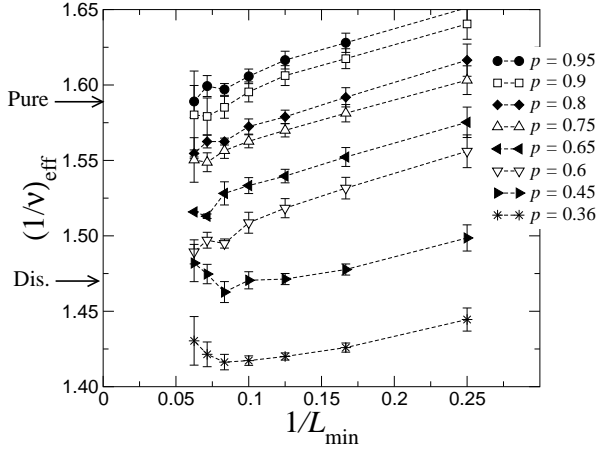}}
\end{center}\vskip -0.3cm
\caption{Effective exponents $(1/\nu)_{{\rm eff}}$ as obtained from
         power-law fits in the range $L_{\rm min} - L_{\rm max}=20$ as 
	 a function of $1/L_{{\rm min}}$ for $p=0.95,\ 0.9,\ 0.8, \ 0.75,\ 0.65,\
         0.6,\ 0.45$, and $0.36$. The error bars correspond to the
         standard deviations of the fits. The arrows indicate
         the values $1/\nu$ for the pure model~\cite{Guida98} and the
         site-diluted one~\cite{Ballesteros98}.
}
\label{1-sur-nu}
\end{figure}

As mentioned in the introduction, the vicinity of the exponent ratios
$\beta/\nu$ and $\gamma/\nu$ for the pure and disordered universality
classes does not allow, by the use of standard finite-size scaling (FSS) 
techniques, to discriminate between
the two fixed points from the behaviour of the magnetisation and the
susceptibility, respectively. Only the critical exponent $\nu$, which can be
evaluated from the asymptotic FSS behaviour of the 
derivative of the magnetisation versus temperature,
\begin{equation}
d\ln\bar m/dK \sim a_{d \ln m} L^{1/\nu},
\label{eq:dmdK}
\end{equation}
will be useful with this technique. But even for $\nu$, one is trying to
resolve an expected shift from the pure model's value by less than $10\%$.
Taking the data evaluated at $K_{\rm max}$ and assuming the leading power-law
behaviour (\ref{eq:dmdK}), we have
extracted by (linear) two-parameter least-squares fits over successively smaller
ranges $L_{{\rm min}}\ -\ L_{{\rm max}}$ the size-dependent effective exponent
$(1/\nu)_{{\rm eff}}$. Using the first set of data for lattice sizes
$L=4, 6, 8, \dots, 20 = L_{\rm max}$, the resulting exponents are plotted
in a broad range of concentrations $p$ against $1/L_{{\rm min}}$
in Fig.~\ref{1-sur-nu} ($L_{{\rm min}}$ is the smallest lattice
size used in the fits). We see that in the regime of low dilution
($p$ close to 1), the system is clearly influenced by the
pure fixed point. On the other hand, when the bond concentration is small, the
vicinity of the percolation fixed point induces a decrease of $1/\nu$ below its
expected disordered value. Indeed, the percolation fixed point is characterized
by $1/\nu\approx 1.12$~\cite{LorenzZiff98}.

\begin{figure}[t]
\begin{center}
        \resizebox{0.80\columnwidth}{!}{%
        \includegraphics{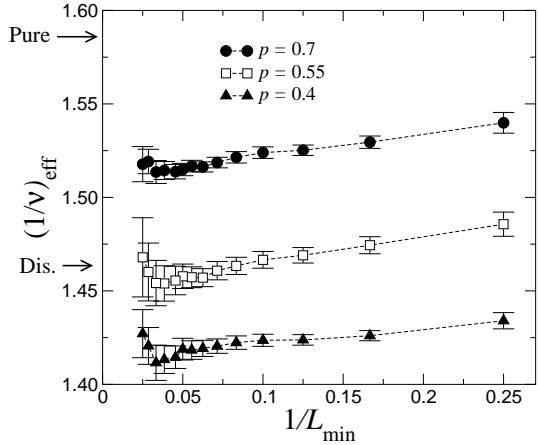}}
\end{center}\vskip -0.3cm
\caption{Effective exponents $(1/\nu)_{{\rm eff}}$ as obtained similar to
Fig.~\ref{1-sur-nu} from the
behaviour of $d \ln \bar m/dK$ as a function of
                 $1/L_{{\rm min}}$ for $p=0.7$, $0.55$, and $0.4$. Here the
upper limit of the fit range is $L_{\rm max} = 96$.}
\label{1-sur-nu-p=0.4-0.55-0.7}
\end{figure}

As is demonstrated in Fig.~\ref{1-sur-nu-p=0.4-0.55-0.7}, the same analysis for
the three mainly studied dilutions with considerably larger lattice
sizes going up to $L=L_{\rm max}=96$ confirms the observation of the great
influence of both the pure and percolation fixed points.
The left-most data points in Fig.~\ref{1-sur-nu-p=0.4-0.55-0.7} with
$L_{\rm min} = 40$ follow from two-parameter least-squares fits
of $(d\ln\bar m/dK)_{K_{\rm max}}$ for our five largest lattice sizes
$L=40, 50, 64, 80$, and $96$. Since this leaves only three independent
degrees of freedom, moving further into the asymptotic large-$L$ regime
would be meaningless with the data available to us. The variation of
$(1/\nu)_{\rm eff}$ with $L_{\rm min}$ clearly indicates deviations of the
data from the asymptotic ansatz (\ref{eq:dmdK}), rooted in confluent
corrections and cross-over terms or both. As will be discussed in more
detail below, we found it impossible, however, to resolve these corrections
within a non-linear four-parameter ansatz. Still, some rough estimates
of the critical exponent $\nu$ can be read off from
Fig.~\ref{1-sur-nu-p=0.4-0.55-0.7}, which are collected in
Table \ref{tab:table2}. Obviously,
it would be very difficult to extract more precise values from
this analysis.
At a qualitative level, however, Fig.~\ref{1-sur-nu-p=0.4-0.55-0.7}
clearly indicates that the dilution for which the cross-over
influence will be the least is around $p=0.55$ which suggests that the scaling
corrections should be rather small for this specific dilution.

\subsubsection{Critical couplings}

\begin{figure}[t]
        \begin{center}
        \resizebox{0.80\columnwidth}{!}{%
        \includegraphics{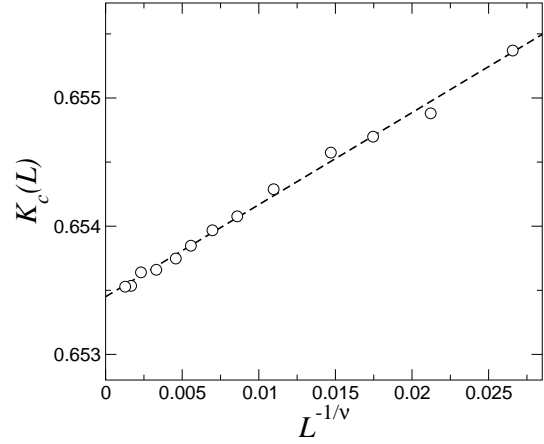}}
        \end{center}\vskip -0.3cm
        \caption{Size-dependent critical coupling $K_c(L)$ versus $L^{-1/\nu}$
                 for the bond-diluted Ising model with $p=0.7$, using our
estimate of $\nu = 0.66(1)$.}
        \label{Kc-vs-1surL}
\end{figure}

The peak locations of the average susceptibility determine with a good accuracy
the size-dependent (pseudo-) critical
couplings $K_c(L)$, and from an infinite-size extrapolation,
$K_c(L) = K_c^\infty + a_K L^{-1/\nu}$, the critical coupling
in the thermodynamic limit can be estimated. Inserting the just determined
estimate of $\nu$, the linear fit shown in Fig.~\ref{Kc-vs-1surL} for $p=0.7$
yields:
\begin{equation}
K_c^\infty =0.6534(1).
\end{equation}

\noindent The same procedure gives for $p=0.55$ and $p=0.4$ the critical
couplings $K_c^\infty=0.8645(2)$ and $K_c^\infty=1.3129(3)$, respectively.

\subsubsection{Susceptibility and magnetisation exponents}

\begin{figure}[b]
\begin{center}
        \resizebox{0.80\columnwidth}{!}{%
        \includegraphics{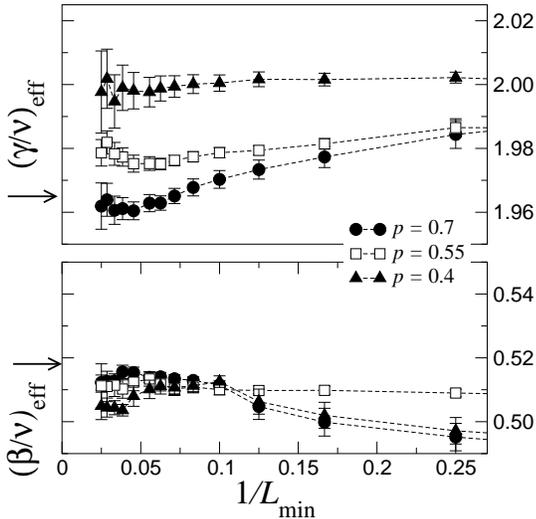}}
\end{center}\vskip -0.3cm
\caption{Effective ratios $(\gamma/\nu)_{\rm eff}$ and
                 $(\beta/\nu)_{\rm eff}$ as a function of $1/L_{\rm min}$ for
                 $p=0.7$, $0.55$, and $0.4$. The error bars correspond to the
                 standard deviations of the power-law fits. The arrows indicate
                 roughly the expected values for both the pure and disordered
                 model.}
\label{exp-effectifs}
\end{figure}

As already mentioned, the average magnetisation $\bar m$ and
susceptibility $\bar\chi$
are expected to scale at the
size-depen\-dent critical coupling $K_c (L)$ with the lattice size as:
\begin{equation}
\bar m_{K_{{\rm max}}}\sim a_m L^{-\beta/\nu}, \quad \bar\chi_{{\rm max}}\sim a_\chi L^{\gamma/\nu},
\end{equation}
where $a_m$ and $a_\chi$ are non-universal amplitudes. From least-squares fits
to our extensive Monte
Carlo data, we computed the effective size-dependent ratios of the
critical exponents $(\beta/\nu)_{\rm eff}$ and $(\gamma/\nu)_{\rm eff}$ which
are plotted in Fig.~\ref{exp-effectifs} against $1/L_{\rm min}$ for
$p=0.7,\ 0.55$, and $0.4$, where $L_{\rm min}$ is again the smallest lattice
size used in the fits.

Concerning the magnetisation, the effective ratios clear\-ly converge,
when the
size increases, towards $0.515(5)$ for the three dilutions in agreement with the
expected values of the pure and site-diluted models. For the susceptibility,
on the other hand, the behaviour of $(\gamma/\nu)_{\rm eff}$ is more
differentiated depending on the concentration $p$: the cases $p=0.7$ and
$0.55$ are compatible with $1.965(10)$ if we take into account the error bars,
very close to the pure and site-diluted model values, but the case $p=0.4$
displays a $(\gamma/\nu)_{\rm eff}$ exponent a little bit larger.
This discrepancy is probably due to the influence of the percolation fixed
point whose $\gamma/\nu$ ratio is close to 2.05, according to the same
scenario as for the exponent $1/\nu$.
The critical exponents from the FSS study are summarized in
Table~\ref{tab:table2} for the three studied dilutions, in good agreement with the
hyperscaling relation $d=2\beta/\nu+\gamma/\nu$.

\begin{table}
\small
\caption{Critical exponents deduced from the FSS study of the three dilutions.}
\vglue0mm\begin{center}
\begin{tabular}{@{}*{5}{l}}
\br
  $p$ & $0.7$   & $0.55$  &  $0.4$   \\
\hline
$1/\nu$                 & $1.52(2)$   & $1.46(2)$    & $1.42(2)$   \\
$\nu$                   & $0.660(10)$ & $0.685(10)$  & $0.705(10)$ \\
$\beta/\nu$             & $0.515(5)$  & $0.513(5)$   & $0.510(5)$  \\
$\gamma/\nu$            & $1.965(10)$ & $1.977(10)$  & $2.000(10)$ \\
$2\beta/\nu+\gamma/\nu$ & $2.995(20)$ & $ 3.003(20)$ & $3.020(20)$ \\
\br
\end{tabular}
\\
\medskip\end{center}
\label{tab:table2}
\end{table}

\begin{figure}[b]
\begin{center}
        \resizebox{0.80\columnwidth}{!}{%
        \includegraphics{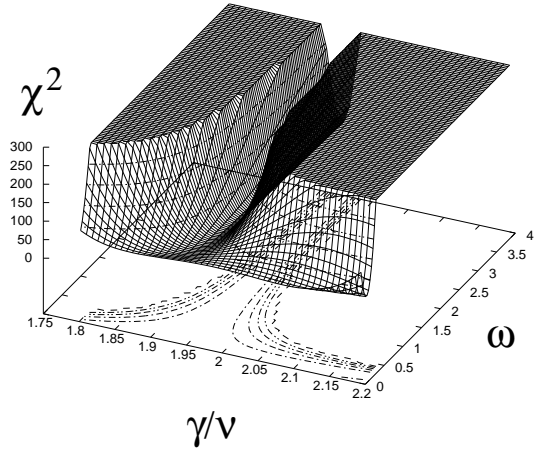}}
\end{center}\vskip 0cm
\caption{Plot of the $\chi^2$-landscape deduced from linear fits of
                 $\bar\chi_{\rm max}(L)=a_\chi L^{\gamma/\nu}(1+b_\chi
                 L^{-\omega})$ in the range $18\leq L\leq 96$ for the
                 bond-diluted Ising model with concentration $p=0.7$,
exhibiting an absolute minimum at
$\gamma/\nu=1.952,\ \omega=2.24$ and a secondary minimum at
$\gamma/\nu=2.112,\ \omega=0.01$. The
                 exponents are fixed parameters and the amplitudes are free.
                 The cutoff at $\chi^2=300$ has been introduced in order to
                 improve the clarity of the figure.}
\label{gsurn}
\end{figure}

Following the work of Ballesteros {\em et al.\/}~\cite{Ballesteros98} which
emphasized the great influence of corrections-to-scaling to get an accurate
determination of the critical exponents in the site-diluted Ising model, we
tried to estimate them in the bond-diluted version as well. If we consider an
irrelevant scaling field $g$ with scaling dimension $y_g=-\omega<0$, the
scaling expression of, e.g., the susceptibility is:
\begin{equation}
\bar\chi (L^{-1},t,g)=L^{\gamma/\nu}f_\chi (Lt^\nu,L^{-\omega}g).
\end{equation}
At $t=0$, around the fixed point value $g=0$, this leads to the standard
expression $a_\chi L^{\gamma/\nu}[1+b_\chi L^{-\omega}+O(L^{-2\omega})]$. We
hence tried to fit the susceptibility data using the ansatz:
\begin{equation}
\bar\chi_{\rm max}(L)=a_\chi L^{\gamma/\nu}(1+b_\chi L^{-\omega}).
\end{equation}
We systematically varied the exponent ratio $\gamma/\nu$ and the
corrections-to-scaling exponent $\omega$
and determined the minimum of $\chi^2/$d.o.f by performing linear fits in
the range $18\leq L\leq 96$ to fix
the amplitudes $a_\chi$ and $b_\chi$. The resulting $\chi^2$-landscape for
$p=0.7$ is shown in Fig.~\ref{gsurn}, where the base plane was restricted to
the range $1.8\leq\gamma/\nu\leq 2.2$ and $0\leq\omega\leq 4$. The
absolute minimum was found at $\gamma/\nu=1.952,\ \omega=2.24$, and
a second, less pronounced minimum at $\gamma/\nu=2.112,\ \omega=0.01$.
The results for the magnetisation respectively its logaritmic derivative
and the other dilutions $p=0.55$ and $0.4$ look qualitatively similar.

The figure may again be interpreted in favor of a competition between the two
fixed points: the first one is characterized by a valley in the
$\omega$-direction, with an absolute minimum for a large value of $\omega$,
i.e., small corrections-to-scaling since these corrections scale as
$L^{-\omega}$.
Then, this fixed point should be the disordered one and the second the
percolation fixed point for which $\gamma/\nu\simeq 2.05$ and
$\beta/\nu\simeq 0.48$~\cite{LorenzZiff98}. But the accurate determination of
the corrections-to-scaling exponent $\omega$ is very difficult because it
strongly depends on the quantity considered ($\bar\chi_{\rm{max}}$,
$\bar m_{K_{\rm{max}}}$ or $(d\ln\bar m/dK)_{K_{\rm{max}}}$) and on the sizes
included in the fits. Indeed,
the valley in the $\omega$ direction does not display any deep minimum and
the absolute minimum observed does not seem to be relevant concerning the
value of $\omega$. We are thus not able to extract a numerical value of the
corrections-to-scaling exponent, in contrast to previous claims that
$\omega \approx 0.4$~\cite{Ballesteros98,Folk99,Folk00}.

\subsection{Temperature Scaling}
\label{s5}

\begin{figure}[t]
\begin{center}
        \resizebox{0.90\columnwidth}{!}{%
        \includegraphics{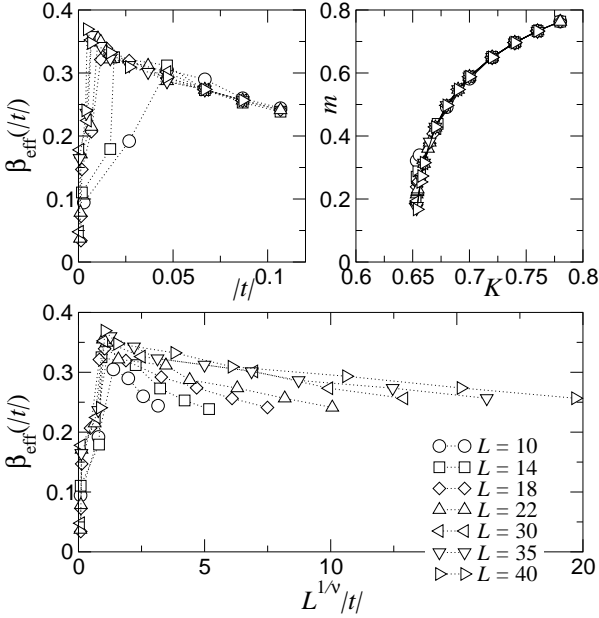}}
\end{center}\vskip 0cm
\caption{Variation of the temperature dependent effective critical exponent
$\beta_{\rm{eff}}(|t|)=d\ln \bar m/d\ln |t|$
(in the low-temperature phase)
as a function of the reduced
temperature $|t|$ (top) and $L^{1/\nu}|t|$ (bottom) for the bond-diluted Ising
model with $p=0.7$ and several lattice sizes
$L$.
The data points are
directly obtained from Monte Carlo data
(no histogram reweighting). The
magnetisation vs.\ the coupling strength
$K=J/k_B T$ in the ordered phase
$K>K_c^\infty$ is shown in the upper part.}
\label{m-beta-p=.7-MC}
\end{figure}

The temperature behaviour of the susceptibility and magnetisation should allow to avoid the previous
difficulties generated by the vicinity of the ratios of the critical exponents. The critical exponents are
indeed the following:

\noindent pure Ising model: $\beta=0.3258(14)$,
$\gamma=1.2396(13)$ ~\cite{Guida98},

\noindent site-diluted model: $\beta=0.3546(28)$,
$\gamma=1.342(10)$ ~\cite{Ballesteros98}.
\begin{figure}[t]
\begin{center}
        \resizebox{0.90\columnwidth}{!}{%
        \includegraphics{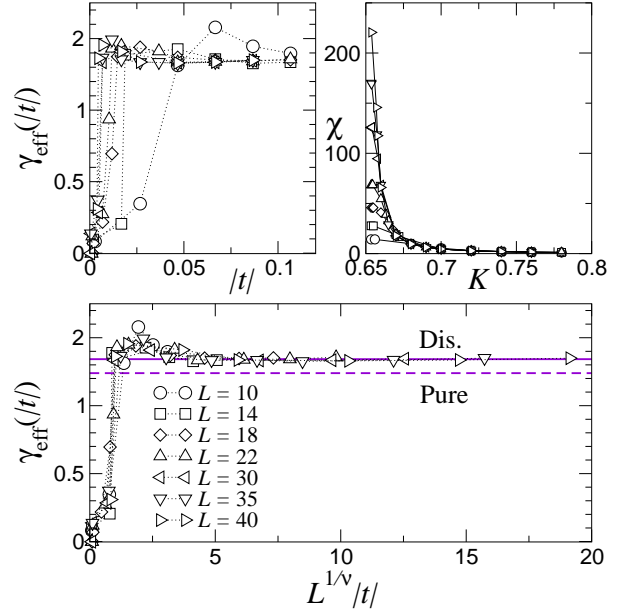}}
\end{center}\vskip 0cm
\caption{Variation of the temperature dependent effective critical exponent
$\gamma_{\rm{eff}}(|t|)=-d\ln \bar\chi/d\ln |t|$
(in the low-temperature phase)
as a function of the reduced
temperature $|t|$ (top) and $L^{1/\nu}|t|$ (bottom)
for the bond-diluted Ising
model with $p=0.7$ and several lattice sizes
$L$.
The horizontal dashed lines indicate
the pure and site-diluted values of $\gamma$.
The
susceptibity vs.\ the coupling strength
$K=J/k_B T$ in the ordered phase is shown in the upper part.}
\label{chi-gamma-p=.7-MC}
\end{figure}

As a function of the reduced temperature $t=(K_c- K)$ ($t<0$ in the
low-temperature (LT) phase and $t>0$ in the high-temperature (HT) phase)
and the system size $L$,
the magnetisation and the susceptibility are expected to scale as:
\begin{equation}
\bar m(t,L)\sim |t|^\beta f_\pm(L^{1/\nu}|t|),
\end{equation}
\begin{equation}
\bar\chi(t,L)\sim |t|^{-\gamma}g_\pm(L^{1/\nu}|t|),
\label{eq-chi-univ}
\end{equation}
where $f_\pm$ and $g_\pm$ are scaling functions of the
variable $x=L^{1/\nu}|t|$ and the subscript $\pm$ stands for the HT/LT phases.
Then we can define temperature dependent effective critical exponents
\begin{eqnarray}
\beta_{\rm{eff}}(|t|)&=&d\ln \bar m/d\ln |t|,\\
\gamma_{\rm{eff}}(|t|)&=&-d\ln \bar\chi/d\ln |t|,
\end{eqnarray}
which should converge towards
the asymptotic critical exponents $\beta$ and $\gamma$ when
$L\rightarrow\infty$ and $|t|\rightarrow 0$. The results for $p=0.7$
are shown in Figs.~\ref{m-beta-p=.7-MC} and~\ref{chi-gamma-p=.7-MC}.

The effective exponents $\beta_{\rm{eff}}(|t|)$ evolve as a function of $|t|$
between a maximum value and 0 in the two $|t|$ directions: when $|t|$ is large
($K$ far from $K_c$), the system is outside the critical region and the
description with the critical exponents is no more valid whereas when
$|t|\rightarrow 0$, the correlation length becomes of the same order as the
linear size of the system and the finite-size effects become very strong.
Nevertheless, just before the curves bend down, for the largest lattice sizes
we can read off that $\beta_{\rm eff}(|t|) \approx 0.34 - 0.36$.
In the case of the susceptibility, for the greatest sizes,
$\gamma_{\rm{eff}}(|t|)$ is stable around $1.34$ when $|t|$ is not too small,
i.e., when the finite-size effects are not too strong. The plot
of $\gamma_{\rm{eff}}(|t|)$ vs.\ the rescaled variable $L^{1/\nu}|t|$
shows that
the critical power-law behaviour holds in different temperature ranges for the
different sizes studied. As expected, the size effects are more sensitive when
the lattice size is small and the critical behaviour is better described when
the size increases.

\begin{figure}[t]
\begin{center}
        \resizebox{0.90\columnwidth}{!}{%
        \includegraphics{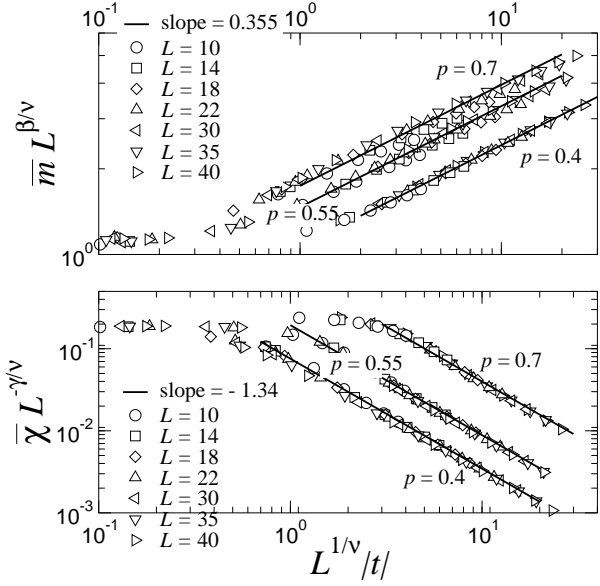}}
\end{center}\vskip 0cm
\caption{Log-log plot of the scaling functions $\bar mL^{\beta/\nu}$ (top) and
$\bar\chi L^{-\gamma/\nu}$ (bottom) against $L^{1/\nu}|t|$
for $p=0.4$, $0.55$, and
$0.7$ in the low-temperature ordered phase ($K>K_c^\infty$). The lines
show the power-law behaviours with the exponents $\beta\simeq 0.355$ and
$\gamma\simeq 1.34$ which characterize the disorder fixed point. The data for
the smallest values of $L^{1/\nu}|t|$
which do not lye on the master curve show
the size effects described previously when the correlation length $\xi$ is
limited by the linear size $L$. }
\label{scaling-m-chi-all_ps-MC}
\end{figure}

A more quantitative analysis is possible for the concentrations $p=0.55$ and
0.7 for which the most accurate temperature simulations have been done. >From the
temperature behaviour of the susceptibility, we have directly extracted the
power-law exponent $\gamma$ from error weighted least-squares fits by
choosing the temperature range that gives the smallest $\chi^2/$d.o.f for
several system sizes. The results are given in Table~\ref{tab:table4}. We see that
all estimates are consistent with $\gamma \approx  1.34 - 1.36$, which is clearly
different from the pure model's exponent of $\gamma \approx 1.24$.
In particular we do not observe in this analysis any pronunced differences
between the two dilutions. For $p=0.4$, our main dilution closest to the
percolation threshold, the number of accurate temperature points is unfortunately
not large enough to allow for reliable fits.
       \begin{table}
       \small
        \caption{Critical exponent $\gamma$ deduced from log-log fits of
the susceptibility vs. the
reduced temperature $|t|$
in the ordered phase
for the concentrations $p=0.55$ and 0.7.}
\vglue0mm\begin{center}
        \begin{tabular}{@{}*{7}{l}}

\hline
$p=0.55$ & $L$                  & 40      &         & 22      &         \\
        \hline
     & $\#$ {\rm points}        & 7       &         & 12      &         \\
     &  $\chi^2/$d.o.f          & 0.07    &         & 1.6     &         \\
     &  $\gamma$    & 1.36(3)& & 1.36(2) &  \\
\hline
$p=0.7$ & $L$                   & 40      & 35      & 22      & 18      \\
        \hline
        & $\#$ {\rm points}     & 10      & 8       & 9       & 8       \\
        & $\chi^2/$d.o.f        & 8.37    & 0.01    & 0.08    & 0.81    \\
        & $\gamma$              & 1.32(1) & 1.34(5) & 1.40(4) & 1.34(5) \\
\hline
\end{tabular}
\\
\medskip\end{center}
\label{tab:table4}
\end{table}

 From the previous expressions of the magnetisation and susceptibility as a
function of the reduced temperature and size, following a procedure
proposed by Binder and Landau~\cite{Binder80} a long time ago, it is instructive
to plot the scaling functions $f_\pm(x)$ and $g_\pm(x)$.
For finite size and $|t|\not=0$, the scaling functions may be Taylor
expanded in powers of the inverse scaling variable
$x^{-1}=(L^{1/\nu}|t|)^{-1}$,
for example in the case of the susceptibility
$\bar \chi_\pm(t,L)=|t|^{-\gamma} [g_\pm(\infty)+
x^{-1}g'_\pm(\infty)+O(x^{-2})]$,
where the amplitude $g_\pm(\infty)$ is usually denoted by $\Gamma_\pm$.
Multiplying by $L^{-\gamma/\nu}$ leads to
\begin{equation}
\bar\chi_\pm L^{-\gamma/\nu}=\tilde g_\pm(x)=\Gamma_\pm x^{-\gamma}+O(x^{-\gamma-1}),
\end{equation}
where $\tilde g_\pm(x) = x^{-\gamma} g_\pm(x)$.
The case of the magnetisation is slightly different, since the magnetisation
is asymptotically vanishing in the high-temperature phase, and thus
$\bar m_\pm(t,L)=|t|^{\beta} [f_\pm(\infty)+
x^{-1}f'_\pm(\infty)+O(x^{-2})]$ or
\begin{equation}
\!\bar m_\pm L^{\beta/\nu}=\tilde f_\pm(x)=x^{\beta}\left\{\matrix{
&\!\!\!\!\!\!0  &\!\!\!+\!\!\!&B'_+ x^{-1} &\!\!\!+\!\!\!&O(x^{-2})\cr
&\!\!\!\!\!\!B_-&\!\!\!+\!\!\!&B'_- x^{-1} &\!\!\!+\!\!\!&O(x^{-2})\cr
}\right.
\end{equation}
which should give universal curves for the different sizes and temperatures.

The curves in the ordered phase shown in Fig.~\ref{scaling-m-chi-all_ps-MC}
are obviously universal master curves whose slopes, in a
log-log plot, give the critical exponents $\beta\simeq 0.355$ and
$\gamma\simeq 1.34$. Indeed, when $|t|\rightarrow 0$ but with $L$ still
larger than the correlation length $\xi$, one should recover the critical
behaviour given by
$\tilde{f}_-(x)\sim x^\beta$ and $\tilde{g}_-(x)\sim x^{-\gamma}$.
The same procedure applied to the two other bond concentrations $p=0.55$ and
$0.4$ gives analogous results and confirms the universal values of the
disordered critical exponents $\beta$ and $\gamma$, independent of
the concentration as is illustrated in Fig.~\ref{scaling-m-chi-all_ps-MC}.

\begin{figure}[t]
\begin{center}
        \resizebox{0.90\columnwidth}{!}{%
        \includegraphics{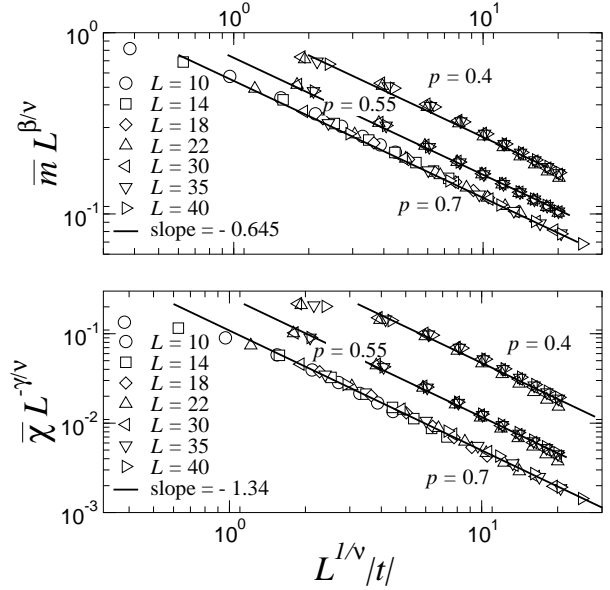}}
\end{center}\vskip 0cm
\caption{Log-log plot of the scaling functions $\bar mL^{\beta/\nu}$ (top)
and $\bar\chi L^{-\gamma/\nu}$ (bottom) against $L^{1/\nu}|t|$
in the disordered
phase ($K<K_c^\infty$) for the three dilutions.
The lines show the power-law behaviours
with the exponents $\beta - 1 \simeq -0.645$ and $-\gamma\simeq -1.34$ which
characterize the disorder fixed point.}
\label{scaling-m-chi-desord-all_ps-MC}
\end{figure}

In the disordered phase $K<K_c^\infty$, $t>0$
the previous scaling assumptions for
$\bar m$ and $\bar\chi$ lead to universal functions
whose slopes on a logarithmic scale are equal to $\beta - 1\simeq -0.645$ and
$-\gamma\simeq -1.34$, see Fig.~\ref{scaling-m-chi-desord-all_ps-MC}.

\begin{figure}[t]
\begin{center}
        \resizebox{0.90\columnwidth}{!}{%
        \includegraphics{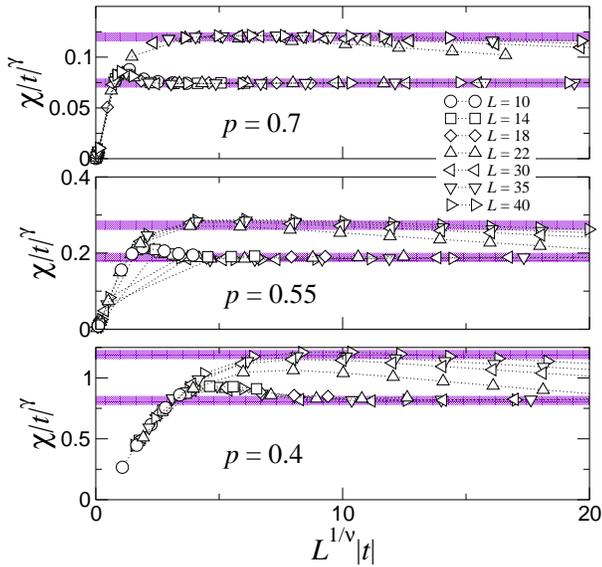}}
\end{center}\vskip 0cm
\caption{Log-log plot of the scaling functions
$\bar\chi |t|^{\gamma}$  against $L^{1/\nu}|t|$ for the three dilutions.
The shaded horizontal stripes indicate the critical amplitudes $\Gamma_+$
and $\Gamma_- (< \Gamma_+)$, respectively.}
\label{chi_temp-rescaled_0.7Both}
\end{figure}

Some combinations of critical amplitudes are also universal and thus
characterize the critical point.
In fact, critical amplitude ratios are often more sensitive
to the universality class than the critical exponents themselves
\cite{Shchur02}.
Among such ratios, $\Gamma_+/\Gamma_-$ is
directly accessible through our results. An example is illustrated in
Fig.~\ref{chi_temp-rescaled_0.7Both}, where the scaling function
of Eq.~(\ref{eq-chi-univ}) is plotted against the scaling variable $x$.
The values of the amplitudes which describe the approach to criticality
from above and from below,
$\Gamma_+$ and $\Gamma_-$, respectively, are shown in shaded horizontal
stripes.
The ratios
\begin{eqnarray}
\Gamma_+/\Gamma_- &=& 1.62\pm 0.10\ (p=0.7),  \label{eq:rat_07} \\
\Gamma_+/\Gamma_- &=& 1.50\pm 0.10\ (p=0.55), \label{eq:rat_055} \\
\Gamma_+/\Gamma_- &=& 1.48\pm 0.20\ (p=0.4)   \label{eq:rat_04}
\end{eqnarray}
follow.
The values obtained for the three dilutions are consistent within error bars
but they unfortunately appear to be in contradiction with a
field-theoretic approach of Bervillier and Shpot~\cite{Bervillier92}, who
predicted a ratio equal to  $\Gamma_+/\Gamma_-=3.05(32)$. This is
nevertheless not a crucially conflicting result, since susceptibility
amplitudes of course depend on the definition used to compute the
susceptibilities.
At any rate, our estimates (\ref{eq:rat_07})--(\ref{eq:rat_04}) for
the disordered Ising model are clearly
different from those for the pure model where $\Gamma_+/\Gamma_-$ obtained
with field-theoretic and high-temperature series expansion techniques varies
between $4.70$ and $4.95$ \cite{Footnote}, and a recent high-precision Monte
Carlo study \cite{Caselle02} concluded that $\Gamma_+/\Gamma_- = 4.75(3)$.

\section{Conclusions}
\label{s6}

In this paper, we have reported an intensive Monte Carlo study of the physical
properties of the 3D disordered bond-diluted Ising model. As the critical
exponents of the pure and disordered models are very close, the numerical
procedure has to be carefully chosen, especially the thermal averaging and the
average over the disorder realisations. We have determined numerically the
phase diagram of the disordered model, in very good agreement with a
single-bond effective-medium approximation. 

The non-self-averaging at
criticality, as discussed by Aharony and Harris, is confirmed here.
At a quantitative level, the discrepancy between the numerical values
of the normalized squared width of the susceptibility
$R_\chi (L)$ in this work and that in Ref.~\cite{Wiseman98b} definitely comes
from the difference in the definition of $\chi$. Although in our case 
$R_\chi (L)$ and to a lesser extent also $R_m(L)$ do not seem to approach 
a unique constant independent 
of the disorder concentration when the system size $L$ increases, we cannot 
exclude a unique limiting value when cross-over effects are properly taken 
into account. This possibility is corroborated by our finite-size scaling
study of critical exponents.
A comparison of the underlying disorder distributions for the bond-diluted
Ising and $q=4$ Potts models shows that the weight of
rare events plays a much more crucial role for the critical properties
of the latter model. 

The main
emphasis was on investigations of the critical behaviour of the disordered
model based on a finite-size scaling study and on a complementary study of
the temperature scaling behaviour. In the first case, we have shown the great
influence of the cross-over phenomena between the pure, disorder and
percolation fixed points which lead to the measurement of effective critical
exponents dependent on the concentration of magnetic bonds. The temperature
study, on the other hand, turned out to be much better behaved and allowed us
to confirm the unique values of the critical exponents, independent of the
concentration and in agreement with the site-diluted values,
thus providing strong numerical evidence for the theoretically predicted
universal behaviour of the two disorder distributions. These results
are summarized in Table~\ref{tab:table5}.
A scaling study of the average of the maxima of the susceptibilities
$\chi_j$ for each sample instead of the maximum of the average susceptibility
would have been an interesting alternative. However, it is technically more
complicated because from sample to sample the position of the maximum of the
susceptibility varies and may be far from the simulation temperature making
histogram reweighting unreliable. There should even appear samples
with large bond-density fluctuations where unconnected or weakly
connected clusters of bonds coexist. Such special configurations
may display a double-peaked (or more complicated) susceptibility curve
resulting from the independence of these uncorrelated clusters.
    \begin{table}
       \small
        \caption{Critical exponents and critical amplitude ratio of the
susceptibility for the different fixed points.}
\vglue0mm\begin{center}
        \begin{tabular}{@{}*{5}{l}}
\br
fixed point & $\nu$ & $\beta$ & $\gamma$ \\
        \hline
       {\rm pure}         & 0.6304(13) & 0.3258(14) & 1.2396(13) \\
       {\rm percolation}  & 0.89       & 0.40       & 1.82       \\
       {\rm site-diluted} & 0.6837(53) & 0.3546(28) & 1.342(10)  \\
       {\rm bond-diluted} & 0.68(2)    & 0.35(1)    & 1.34(1)    \\
\br
fixed point & $\beta/\nu$ & $\gamma/\nu$ & $\Gamma_+/\Gamma_-$ \\
\hline
       {\rm pure}         & 0.517(3) & 1.966(6) & 4.75(3) \\
       {\rm percolation}  & 0.45     & 2.05     &         \\
       {\rm site-diluted} & 0.519(3) & 1.963(5) &         \\
       {\rm bond-diluted} & 0.515(5) & 1.97(2)  & 1.50(20)\\
\br
\end{tabular}
\\
\medskip\end{center}
\label{tab:table5}
\end{table}

Finally,
we should remark that even though this work required a really huge amount of
effort and CPU time, for some reasons that we are unable to identify, the
accuracy of our results is not as good as in the work of Ballesteros {\em et
al.\/}~\cite{BallesterosEtAl00} for site-dilution. For example we were not
able to conclude for a reliable estimate of the corrections-to-scaling exponent
$\omega$ as they did.
Nevertheless it is worth noticing that the temperature scaling is quite
satisfying and led to the measurement of the susceptibility amplitude ratio
$\Gamma_+/\Gamma_-$, a universal combination which was still unknown
numerically, and which complements an estimate of other universal
combinations of amplitudes reported recently in
Refs.~\cite{CalabreseEtAl03Bis,CalabreseEtAl03}. As anticipated from
a previous analysis for disordered systems, our estimates for the amplitude
ratio $\Gamma_+/\Gamma_-$ are clearly different from the pure model and thus
yield the cleanest signal for a unique disorder fixed point.

\begin{acknowledgement}
The authors are happy to thank Yu. Holovatch and
L.N. Shchur for many discussions.
We also gratefully acknowledge financial support by the DAAD and A.P.A.P.E.
through the PROCOPE exchange programme. This work was supported by the
computer-time grants No.\ 2000007 of the Centre
de  Ressources Informatiques de Haute Normandie (CRIHAN), Rouen,
No.\ 062 0011 of CINES, Montpellier,
No.\ hlz061 of NIC, J\"ulich,
and
No.\ h0611 of LRZ, M\"unchen. W.J. also acknowledges partial support
by the EU-Network HPRN-CT-1999-000161 ``Discrete Random Geometries: From
Solid State Physics to Quantum Gravity'' and the German-Israel Foundation
(GIF) under grant No.\ I-653-181.14/1999.
\end{acknowledgement}


\end{document}